# Retrieval over Reasoning: A Cost-Controlled Benchmark of Language Models for Energy-Retrofit Recommendation

**Eliseo Curcio**

## Abstract

Recommending the correct set of energy conservation measures (ECMs) for a building is a structured, multi-label prediction problem in which a task-specific supervised model has weak training signal and a general language model has no grounding in the local building stock. We study this problem on 10,422 real New York City Local Law 87 (LL87) energy-audit records, taking as ground truth the set of ECM categories that certified auditors actually recommended. We make four contributions. First, we establish that energy-use-intensity (EUI) prediction the upstream task is effectively solved by tree ensembles: across fifteen trained models, a stacking ensemble reaches a coefficient of determination $R^2 = 0.757$, and every one of six neural architectures is outperformed by gradient-boosted trees. Second, we show that the framing of the recommendation task dominates model choice: recasting ECM recommendation as 19-way multi-label classification rather than single-label categorization lifts a gradient-boosted-tree baseline from a previously reported 25.9% accuracy to a micro-F1 of 0.571. Third, we benchmark eight large language models (LLMs) from four providers in a 2×2 design that independently toggles retrieval grounding and explicit reasoning, scoring each arm on per-label F1, U.S.-dollar cost per building, and latency; retrieval-augmented generation (RAG) improves micro-F1 by +0.11 to +0.20 on every model, while explicit reasoning yields no measurable accuracy change (−0.018 to +0.010) at up to 8.4× the cost. Fourth, we show LLMs systematically over-recommend high recall, low precision and that retrieval closes the gap chiefly by improving precision. A 70-billion-parameter open-weight model with a fifteen-line nearest-neighbor retrieval step reaches 0.511 micro-F1 at $0.00032 per building, comparable to a frontier model at roughly 10.1× lower cost.

## 1. Introduction

Buildings account for roughly 40% of primary energy use in developed economies, and deciding which retrofit measures a given building requires is a precondition for any large-scale efficiency program [1]. In New York City this judgment is produced by certified energy auditors under Local Law 87 (LL87), which, together with the Local Law 84 (LL84) benchmarking mandate, requires large buildings to disclose energy use and undergo periodic audits [2]. Auditing is costly and slow, so a system that could reproduce auditor recommendations from already-available building characteristics would let program managers triage a portfolio before committing inspection budgets. Prior data-driven work on this corpus has focused on predicting energy use and assigning performance grades [3] [4], and broad reviews confirm that machine learning is now the dominant approach to building-energy estimation [5] [6].

ECM recommendation is intrinsically a structured-prediction problem: a building does not receive a single fix but a *set* of measures. We make three observations. First, the upstream EUI-prediction task is, on this data, effectively solved by tree ensembles such as XGBoost [7], LightGBM [8], and CatBoost [9]; the open problem is recommending the measures themselves. Second, the framing of the recommendation task matters more than the model: a prior single-label, twelve-class formulation reported only 25.9% accuracy, whereas multi-label classification one binary decision per ECM category lifts the same supervised family to a micro-F1 of 0.571. Third, general LLMs are attractive because they require no task-specific training, but without grounding they cannot know what auditors in this stock recommend; retrieval-augmented generation (RAG) [10] supplies that grounding, while the value of explicit reasoning [11] on an associative task is unclear.

We therefore design a cost-controlled benchmark separating these factors. Holding the task, the input representation, and the scoring fixed, we evaluate eight LLMs in a 2×2 grid that independently turns retrieval and reasoning on and off, alongside a supervised gradient-boosted-tree baseline, and we measure dollar cost and latency in addition to accuracy.

**Contributions.** (1) A characterization of the upstream EUI task across fifteen models, establishing tree-ensemble dominance. (2) A reusable benchmark for structured ECM recommendation grounded in 10,422 real auditor records, with an identical input/output contract across supervised and LLM systems. (3) A factorial separation of retrieval from reasoning, establishing retrieval as the dominant lever and reasoning as wasted computation for this task. (4) A cost–accuracy analysis showing model identity is largely irrelevant once retrieval is present.

## 2. Related Work

**Machine learning for building energy.** Data-driven models have largely displaced physics-based simulation for building-energy estimation [5] [6]. On NYC benchmarking data specifically, Kontokosta and Tull built a city-scale predictive model of building energy use [3], and Papadopoulos and Kontokosta graded buildings on performance using the same disclosure data [4]. These efforts target energy *use*; ours targets the *measures* a building needs.

**Tree ensembles and deep tabular models.** Gradient-boosted decision trees [7] [8] [9], descended from random forests [12], are the standard strong baseline for tabular data. Deep architectures residual networks [13], recurrent LSTM networks [14], Transformers [15], and attention-based tabular models such as TabNet [16] and probabilistic Gaussian-process models [17] have all been proposed, yet recent systematic comparison finds tree ensembles still outperform deep learning on typical tabular problems [18]. Our EUI results (Section 4) reproduce this conclusion on auditor data, using standard implementations [19].

**Multi-label classification.** Predicting a set of labels rather than one is the multi-label setting [20] [21]; binary relevance trains one classifier per label and is a strong, simple baseline. Our supervised arm is binary relevance over gradient-boosted trees, and our retrieval arm rests on nearest-neighbor classification [22].

**Language models, retrieval, and reasoning.** Pretrained Transformers [23] and large autoregressive models [24] perform tasks from natural-language prompts without task-specific training. RAG couples a generator with a retrieval store so outputs can be grounded in retrieved evidence [10] [25], and the choice of in-context examples strongly affects quality [26]. Chain-of-thought prompting [11], self-consistency [27], and reason-and-act scaffolds [28] improve performance on compositional tasks; whether such reasoning helps associative recommendation is the question we test. Recursive Language Models (RLMs) [29] handle inputs exceeding the context window, an advantage that does not apply to single-building inputs (Section 8). Holistic and adversarial evaluation frameworks [30] [31] motivate our cost-aware, ground-truth-anchored scoring; interpretability methods such as SHAP [32] informed the upstream feature analysis.

## 3. Problem Formulation

Let each building be a feature vector $x \in \mathbb{R}^d$, and let $\mathcal{L} = \{\ell_1, \ldots, \ell_m\}$ be the m = 19 ECM categories. The ground-truth recommendation is a subset $Y \subseteq \mathcal{L}$, encoded as $y \in \{0,1\}^m$ with $y_j = 1$ iff the auditor recommended $\ell_j$. A system outputs $\hat{y} \in \{0,1\}^m$. Under binary relevance the multi-label problem decomposes into m independent binary decisions, and the supervised objective minimizes the summed binary cross-entropy:

$$\mathscr{L}(\theta) = -\sum_{i=1}^{n} \sum_{j=1}^{m} [\, y_{ij} \log \hat{y}_{ij} + (1 - y_{ij}) \log(1 - \hat{y}_{ij}) \,] \tag{1}$$

Predictions are scored by precision, recall, and their harmonic mean. With TP, FP, FN the true positives, false positives, and false negatives for a label,

$$P = \frac{TP}{TP + FP}, \quad R = \frac{TP}{TP + FN}, \quad F_1 = \frac{2\,P\,R}{P + R} \tag{2}$$

Micro-averaging pools counts across labels before computing $F_1$ (weighting each decision equally, favoring frequent categories); macro-averaging computes per-label $F_1$ then averages (weighting each category equally, exposing rare ECMs):

$$F_1^{\text{micro}} = \frac{2\,\Sigma_j\, TP_j}{2\,\Sigma_j\, TP_j + \Sigma_j\, FP_j + \Sigma_j\, FN_j}, \quad F_1^{\text{macro}} = \frac{1}{m}\sum_{j=1}^{m} F_1^{(j)} \tag{3}$$

For the upstream regression task we report the coefficient of determination, with $\hat{y}_i$ the predicted and $\bar{y}$ the mean EUI:

$$R^2 = 1 - \frac{\Sigma_i\,(y_i - \hat{y}_i)^2}{\Sigma_i\,(y_i - \bar{y})^2} \tag{4}$$

Finally, the dollar cost of one building's LLM prediction, with token prices $c_{\text{in}}, c_{\text{out}}$ and measured token counts $t_{\text{in}}, t_{\text{out}}$, is

$$Cost = c_{\text{in}}\, t_{\text{in}} + c_{\text{out}}\, t_{\text{out}} \tag{5}$$

We report mean cost and mean wall-clock latency per building. We do not report energy in Joules: hosted models run on provider GPUs we cannot instrument, so any energy figure would be an unverifiable estimate.

## 4. Upstream Task: Energy-Use-Intensity Prediction

### 4.1 Purpose and motivation

Before we address the central problem of recommending measures, we characterize the upstream task that any recommendation system implicitly depends on: predicting a building's energy-use intensity (EUI), the annual energy consumed per unit floor area. We do this for two reasons. First, EUI prediction is the canonical supervised problem on this kind of benchmarking data, and establishing how well it can be solved and by which model family tells us what class of model is appropriate when we later build the supervised baseline for recommendation. Second, the comparison answers a question that recurs throughout the paper: on structured building data, does model sophistication help, or is the limiting factor the information available in the inputs? The answer here (model sophistication does not help; trees beat neural networks) directly motivates our later finding that, for recommendation too, the binding constraint is information supplied by retrieval rather than model capacity.

## 4.2 Models compared and why each is included

We trained fifteen models spanning three families, chosen to represent the methods a practitioner would realistically consider for tabular building data. The **tree-ensemble family** comprises gradient-boosted decision trees XGBoost [7], LightGBM [8], and CatBoost [9] together with a random-forest [12] lineage and a stacking ensemble that combines them; these are the standard strong baselines for tabular regression and learn axis-aligned splits over the raw features. The **neural family** comprises six architectures included to test whether learned representations beat trees: a deep multilayer perceptron, a deep residual network [13], a one-dimensional convolutional network, a recurrent LSTM [14], a sequence Transformer [15], and TabNet [16], an attention-based model designed specifically for tabular data. The **probabilistic family** is represented by a Gaussian process [17], included as a calibrated-uncertainty baseline. All models were trained with standard implementations [19] under two formulations: a *panel* formulation (14,332 building-years, in which each building's prior-year EUI is available as a feature) and a *cross-sectional* formulation (46,641 buildings, predicting from physical and use characteristics alone, with no temporal history).

## 4.3 Results and what they mean

Table 1 reports all fifteen models, sorted by panel $R^2$. Reading the table from the top: the stacking ensemble leads, the three gradient-boosted-tree models form the next tier, and critically every one of the six neural architectures falls below them. The pattern is consistent and worth stating plainly before interpreting it.

| Model | Type | Data | $R^2$ | $R^2$ sd | RMSE | MAE |
|---|---|---|---|---|---|---|
| **Ensemble Stack** | Stacking | Panel | **0.757** | ±0.037 | 42.3 | 16.8 |
| LightGBM | Tree | Panel | 0.739 | ±0.044 | 43.8 | 17.7 |
| CatBoost | Tree | Panel | 0.736 | ±0.038 | 44.0 | 16.9 |
| DeepResNet | Neural Net | Panel | 0.725 | ±0.034 | 44.9 | 18.9 |
| XGBoost | Tree | Panel | 0.720 | ±0.035 | 45.3 | 17.7 |
| DeepMLP | Neural Net | Panel | 0.688 | ±0.072 | 47.5 | 18.7 |
| 1D-CNN | Neural Net | Panel | 0.675 | ±0.065 | 48.6 | 18.6 |
| Seq LSTM | Neural Net | Panel | 0.668 | ±0.044 | 49.1 | 18.5 |
| ResNet | Neural Net | Panel | 0.628 | ±0.151 | 51.0 | 19.2 |
| Seq Transformer | Neural Net | Panel | 0.625 | ±0.043 | 52.4 | 23.5 |
| TabNet | Neural Net | Panel | 0.589 | ±0.139 | 54.3 | 20.3 |
| Gaussian Process | Probabilistic | Panel | 0.513 | — | 69.3 | 24.4 |
| XGB Tuned (cross) | Tree | Cross-sect | 0.463 | ±0.032 | 67.7 | 29.6 |
| CatBoost (cross) | Tree | Cross-sect | 0.421 | ±0.032 | 70.3 | 31.9 |
| LightGBM (cross) | Tree | Cross-sect | 0.408 | ±0.033 | 71.0 | 32.9 |

*Table 1. EUI prediction across fifteen models, sorted by panel $R^2$. Panel models use 14,332 building-years (prior-year EUI available); cross-sectional models use 46,641 buildings (no temporal history). $R^2$ standard deviation from cross-validation where available. The stacking ensemble leads; all six neural architectures trail the gradient-boosted trees.*

Two facts are decisive. First, the stacking ensemble attains $R^2 = 0.757$ on the panel data, and the three gradient-boosted-tree models (LightGBM 0.739, CatBoost 0.736, XGBoost 0.720) occupy the next tier.

Second, every one of the six neural architectures deep MLP, deep residual network, 1D-CNN, sequence LSTM, sequence Transformer, and TabNet falls below the best trees, with the most expressive sequence models (Transformer 0.625, TabNet 0.589) among the weakest of the deep set, and the Gaussian process lowest at 0.513. The cross-sectional models, lacking prior-year EUI, plateau near $R^2 = 0.463$, confirming that temporal history carries most of the predictable signal. This pattern tree ensembles outperforming deep networks on tabular data matches the broader tabular-learning literature [18], and it justifies using gradient-boosted trees, not a neural network, as the supervised baseline for the recommendation task. For practical purposes the EUI task is solved; the difficulty lies downstream.

The gap between the panel and cross-sectional formulations is itself informative. The panel models reach $R^2$ around 0.72–0.76 because each building's prior-year EUI is available as a feature, and a building's energy intensity is highly autocorrelated year over year, so knowing last year's value carries most of the way to predicting this year's. The cross-sectional models, which must predict from physical and use characteristics alone, plateau near 0.46 still useful for ranking but far less precise. The practical reading is that EUI prediction is largely a problem of exploiting history, not of architectural sophistication, which is exactly why the simplest strong tabular learner wins and why the deep sequence models, built to extract structure the tabular features do not contain, gain nothing. We carry one lesson forward: on this data, model capacity is not the binding constraint the binding constraint is whether the relevant signal is present in the input at all. That theme returns, transformed, when we find that retrieval rather than model scale drives recommendation quality.

## 5. Data

Having established that the upstream EUI task is solved by tree ensembles and limited by input information rather than model capacity, we now turn to the central problem: recommending the measures themselves. Unlike EUI, this task has no continuous target and no obvious single-model solution it requires predicting a set of expert decisions so we describe its data and formulation carefully before introducing the methods that attack it.

### 5.1 Source and ground truth

Our ground truth is the LL87 energy-audit dataset for 2019–2024, comprising 10,702 building audit records [2]. Each record lists, across up to three measure packages and up to eight measures per package, the ECM category an auditor recommended, with measure-level cost and expected-life fields. Building characteristics borough, gross and conditioned floor area, above- and below-grade floor counts, space-function use types, occupancy, dwelling-unit counts, and total annual energy use and cost are drawn from the audit records and the matched LL84 benchmarking dataset. After removing records with no usable category, 10,422 buildings remain.

### 5.2 Label space and cardinality

We take each building's label as the union of recommended ECM categories across its measure slots, discarding non-actionable bins, and retain the 19 categories occurring in at least 50 audits so that each binary label has adequate support. The label cardinality is 2.8 categories per building on average (median 3); Table 2 gives the distribution. That most buildings receive two to four measures is the empirical reason the prior single-label formulation was capacity-limited: it was forced to choose one answer where auditors give several.

| Categories per building | Buildings | Share |
|---|---|---|
| 1 | 1,031 | 9.9% |
| 2 | 3,017 | 28.9% |
| 3 | 3,866 | 37.1% |
| 4 | 2,373 | 22.8% |
| 5 or more | 135 | 1.3% |

*Table 2. Distribution of recommended ECM categories per building (n = 10,422). Mean 2.8, median 3.*

### 5.3 Split and leakage control

We use an 80/20 split: 8,337 buildings for training and retrieval, 2,085 held out. The supervised baseline is scored on the full held-out set; the LLM arms, for cost reasons, on a fixed random sample of 200 held-out buildings, identical across arms. **Leakage control:** retrieval neighbors are drawn only from the training split, and a test building's own labels are never visible to any system [22].

## 6. Method

This section describes exactly what the benchmark does, end to end, so the procedure can be reproduced. The pipeline has five stages. (1) We load the 10,422 labeled audit records and construct, for each building, a feature vector from its physical and energy characteristics and a binary label vector over the 19 ECM categories. (2) We split the buildings 80/20 into a training pool and a held-out test set, fixing the random seed so the split is identical across every arm. (3) We fit the supervised baseline on the training pool, and we build the retrieval index over the same pool. (4) For each held-out building we run all five arms, each of which takes the identical feature representation and must emit the identical object a subset of the 19 categories. (5) We compare each emitted subset against the auditor's real recommendation and record accuracy, dollar cost, and latency. The only thing that varies across arms is how the prediction is produced; the inputs, the label space, and the scoring are held constant. That control is what licenses attributing differences in the results to the architecture rather than to incidental differences in setup.

All five arms consume the same building feature representation and emit the same object a subset of the 19 ECM categories. The arms differ only in how they reach the prediction; this symmetry makes the comparison valid.

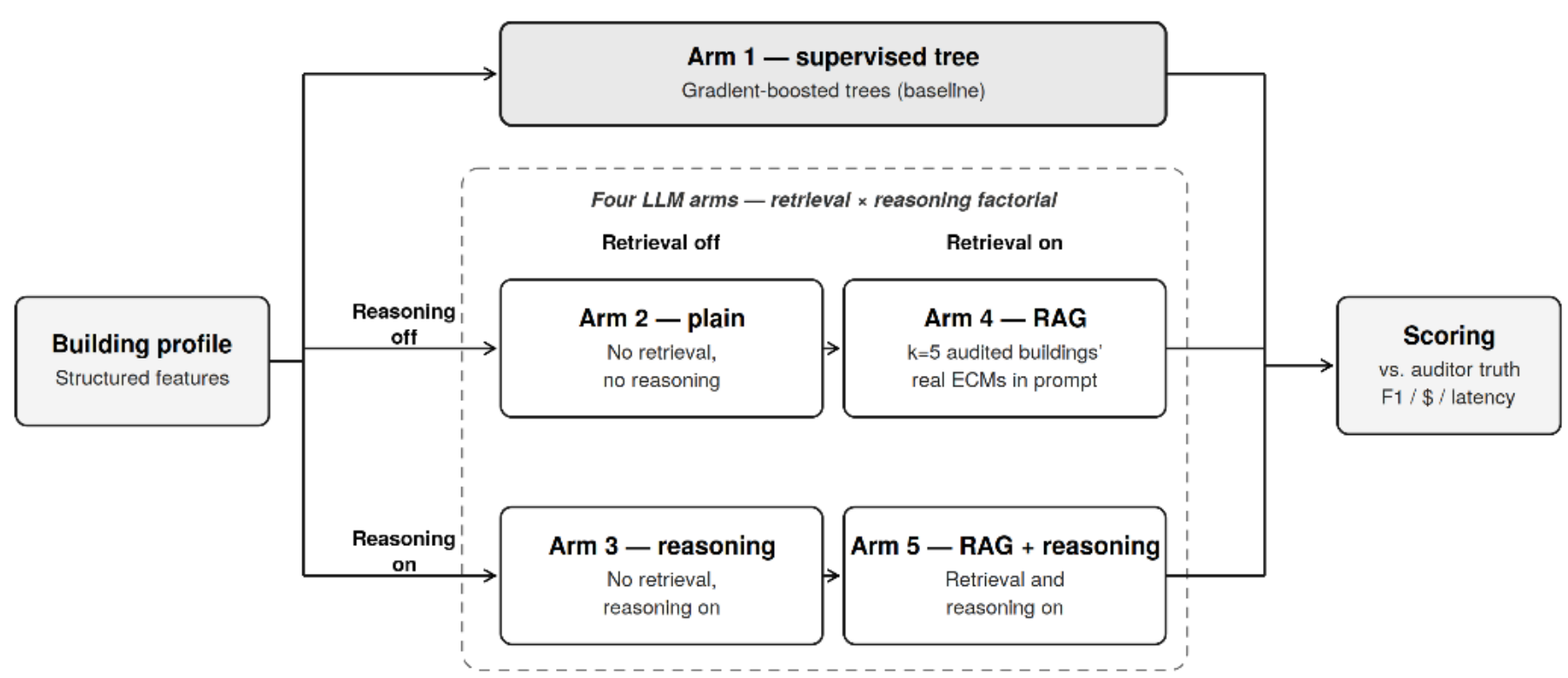

*__Figure 1.__ Five-arm benchmark comparing a supervised tree baseline against four LLM configurations in a retrieval × reasoning 2×2 factorial. All arms share identical input and output; only the prediction method varies. Predicted ECM sets are scored against auditor ground truth (micro/macro-F1, $ per building, latency).*

### 6.1 Arms

**Arm 1: Supervised tree (baseline).** Binary-relevance gradient-boosted trees [7], one head per ECM category, trained on the 8,337-building training split. Section 4 justifies this choice over a neural network.

**Arm 2: Plain LLM.** The building profile and the allowable categories are given to the model [24], which returns a category subset as JSON. No retrieval, no reasoning.

**Arm 3: Reasoning LLM.** As Arm 2 with explicit reasoning enabled (native reasoning-effort or thinking mode where supported, else a step-by-step instruction) [11].

**Arm 4: RAG LLM.** We retrieve the k = 5 nearest training buildings under Euclidean distance on standardized features [22] [10], and insert their real auditor ECM sets into the prompt as evidence [26]:

$$N(x) = arg\,min_{(k)} \parallel z(x) - z(x') \parallel_2, \; x' \in train \qquad (6)$$

where z(·) is standardization. Retrieval is a nearest-neighbor index of roughly fifteen lines of code; no vector database is required. No reasoning.

**Arm 5: RAG + Reasoning.** Retrieval and reasoning both enabled: the fourth corner of the 2×2.

### 6.2 Models and scoring

We evaluate eight models across four providers (Table 3): three Anthropic Claude models (Opus 4.8, Sonnet 4.6, Haiku 4.5), two OpenAI models (GPT-5.5, GPT-5-mini), Google Gemini 3.5 Flash, and two open-weight models (Llama 3.3 70B Instruct, Qwen3 235B) served via a hosted API. Each runs through all four LLM arms, giving 32 model–arm configurations plus the supervised baseline. Scoring uses Eq. 3 for accuracy, Eq. 5 for cost from measured tokens at current list prices, and mean wall-clock latency.

| Model | Provider | Class | Access |
|---|---|---|---|
| Claude Opus 4.8 | Anthropic | Frontier | API |
| Claude Sonnet 4.6 | Anthropic | Mid | API |
| Claude Haiku 4.5 | Anthropic | Small | API |
| GPT-5.5 | OpenAI | Frontier | API |
| GPT-5-mini | OpenAI | Small | API |
| Gemini 3.5 Flash | Google | Mid | API |
| Llama 3.3 70B Instruct | Meta | Open-weight | Hosted |
| Qwen3 235B | Alibaba | Open-weight | Hosted |

*Table 3. The eight evaluated LLMs.*

## 7. Results

We report results at three levels of granularity, each answering a different question. The architecture summary (Section 7.1) answers “which approach is best on average?” The per-model deltas (Section 7.2) answer “does the conclusion hold for every model, or only on average?” a stronger test, because an effect that appears only in the mean may be an artifact of one model. The full scoreboard (Section 7.3) gives every configuration for readers who want the underlying numbers, and the cost–accuracy frontier (Section 7.4) answers the deployment question of accuracy per dollar. Throughout, the supervised tree is the reference line the LLM arms are measured against, and we read accuracy together with cost, because an arm that is marginally more accurate at several times the price is not obviously preferable.

## 7.1 Architecture summary

Table 4 averages each architecture across all eight models. Reading the rows in order: the plain LLM reaches micro-F1 0.33, adding reasoning leaves it essentially unchanged at 0.33, adding retrieval lifts it to 0.49, and adding both holds at 0.48. The supervised tree leads all LLM arms at 0.571, but does so with high precision (0.755) and low recall (0.459) the opposite balance from the LLMs, whose recall exceeds 0.74 throughout. The cost and latency columns show the second theme: the two retrieval arms are no more expensive than plain on average, whereas the reasoning arms cost more and run slower.

| Architecture | Micro-F1 | Macro-F1 | Prec. | Rec. | $/bldg | s/bldg |
|---|---|---|---|---|---|---|
| **Supervised tree (baseline)** | **0.571** | 0.484 | 0.755 | 0.459 | ~0 | 0.01 |
| Plain LLM | 0.328 | 0.220 | 0.206 | 0.813 | $0.00197 | 1.79 |
| Reasoning LLM | 0.325 | 0.219 | 0.206 | 0.798 | $0.00446 | 2.62 |
| RAG LLM | 0.494 | 0.318 | 0.373 | 0.750 | $0.00176 | 1.23 |
| RAG + Reasoning | 0.481 | 0.303 | 0.368 | 0.746 | $0.00369 | 2.14 |

*Table 4. Per-architecture results averaged across all eight LLMs (n = 200 per arm). Tree inference cost and latency are effectively zero.*

## 7.2 Retrieval is the dominant lever

Table 5 isolates each switch per model as the change in micro-F1 against the plain prompt. The retrieval column is positive for all eight models and never below +0.107; the reasoning column straddles zero, from −0.018 to +0.010. The final column shows what reasoning costs: on the small reasoning-heavy models it multiplies per-building cost by 4.1× (GPT-5.5), 5.9× (Gemini), and 8.4× (GPT-5-mini), with no accuracy return. The conclusion is unambiguous at the level of individual models, not merely on average: retrieval helps everywhere, reasoning helps nowhere.

| Model | Retrieval Δ F1 | Reasoning Δ F1 | Reason cost × |
|---|---|---|---|
| Claude Opus 4.8 | **+0.170** | +0.011 | 1.0× |
| Claude Sonnet 4.6 | **+0.172** | -0.013 | 1.1× |
| Claude Haiku 4.5 | **+0.146** | +0.000 | 1.0× |
| GPT-5.5 | **+0.204** | +0.006 | 4.3× |
| GPT-5-mini | **+0.106** | -0.019 | 8.5× |
| Gemini 3.5 Flash | **+0.158** | -0.004 | 5.8× |
| Llama 3.3 70B | **+0.197** | +0.002 | 1.0× |
| Qwen3 235B | **+0.172** | -0.006 | 0.8× |

*Table 5. Effect of retrieval and reasoning per model, as change in micro-F1 vs. plain. "Reason cost ×" is the reasoning arm's per-building cost multiple relative to plain.*

## 7.3 Full scoreboard

Table 6 reports every configuration. The RAG rows (bold) cluster between 0.40 and 0.52 micro-F1 across all eight models an 0.12-wide band whereas the plain rows span a similar band at a markedly lower level.

Once retrieval is present, the spread attributable to model identity is small relative to the gain from retrieval itself.

| Model / Arm | F1 | Prec | Rec | MacF1 | $/bldg | s |
|---|---|---|---|---|---|---|
| **Tree (supervised)** | **0.571** | 0.755 | 0.459 | 0.484 | ~0 | 0.01 |
| Claude Opus 4.8: plain | 0.350 | 0.221 | 0.840 | 0.217 | $0.00596 | 0.4 |
| reason | 0.361 | 0.231 | 0.829 | 0.222 | $0.00607 | 0.4 |
| RAG | **0.520** | 0.411 | 0.710 | 0.317 | $0.00571 | 0.3 |
| RAG+rea | 0.528 | 0.427 | 0.690 | 0.317 | $0.00581 | 0.3 |
| Claude Sonnet 4.6: plain | 0.330 | 0.205 | 0.836 | 0.223 | $0.00248 | 0.4 |
| reason | 0.317 | 0.195 | 0.847 | 0.225 | $0.00266 | 0.3 |
| RAG | **0.502** | 0.374 | 0.762 | 0.321 | $0.00229 | 0.3 |
| RAG+rea | 0.476 | 0.340 | 0.790 | 0.301 | $0.00244 | 0.3 |
| Claude Haiku 4.5: plain | 0.354 | 0.223 | 0.854 | 0.235 | $0.00097 | 0.2 |
| reason | 0.353 | 0.222 | 0.861 | 0.227 | $0.00100 | 0.2 |
| RAG | **0.500** | 0.376 | 0.746 | 0.314 | $0.00081 | 0.2 |
| RAG+rea | 0.488 | 0.359 | 0.763 | 0.314 | $0.00085 | 0.2 |
| GPT-5.5: plain | 0.322 | 0.202 | 0.795 | 0.220 | $0.00358 | 0.3 |
| reason | 0.327 | 0.202 | 0.865 | 0.224 | $0.01534 | 1.7 |
| RAG | **0.526** | 0.407 | 0.742 | 0.345 | $0.00318 | 0.6 |
| RAG+rea | 0.527 | 0.424 | 0.696 | 0.324 | $0.01245 | 1.2 |
| GPT-5-mini: plain | 0.295 | 0.177 | 0.890 | 0.229 | $0.00027 | 0.4 |
| reason | 0.276 | 0.160 | 0.998 | 0.232 | $0.00229 | 2.0 |
| RAG | **0.401** | 0.258 | 0.902 | 0.310 | $0.00024 | 0.3 |
| RAG+rea | 0.325 | 0.196 | 0.950 | 0.237 | $0.00192 | 1.6 |
| Gemini 3.5 Flash: plain | 0.350 | 0.230 | 0.740 | 0.205 | $0.00127 | 2.5 |
| reason | 0.347 | 0.242 | 0.616 | 0.208 | $0.00734 | 8.6 |
| RAG | **0.509** | 0.405 | 0.685 | 0.306 | $0.00109 | 3.9 |
| RAG+rea | 0.518 | 0.449 | 0.612 | 0.313 | $0.00516 | 5.7 |
| Llama 3.3 70B: plain | 0.314 | 0.197 | 0.774 | 0.206 | $0.00026 | 0.1 |
| reason | 0.316 | 0.205 | 0.698 | 0.200 | $0.00027 | 0.1 |
| RAG | **0.511** | 0.387 | 0.753 | 0.324 | $0.00032 | 0.1 |
| RAG+rea | 0.502 | 0.374 | 0.765 | 0.302 | $0.00034 | 0.1 |
| Qwen3 235B: plain | 0.308 | 0.192 | 0.774 | 0.222 | $0.00096 | 10.1 |
| reason | 0.302 | 0.195 | 0.673 | 0.213 | $0.00073 | 7.7 |
| RAG | **0.480** | 0.365 | 0.701 | 0.310 | $0.00040 | 4.2 |
| RAG+rea | 0.486 | 0.373 | 0.698 | 0.313 | $0.00058 | 7.7 |

*Table 6. Full scoreboard: eight models × four LLM arms plus the supervised tree (n = 200 per arm; Gemini arms scored on 168–199 buildings due to transient provider errors). RAG rows bold.*

To confirm these differences are not artifacts of the 200-building sample, we computed bootstrap 95% confidence intervals by resampling the held-out buildings with replacement (2,000 iterations), pooling label-level counts within each resample as in the primary scoring. The intervals are: plain 0.327 [0.312, 0.341], reasoning 0.325 [0.310, 0.340], RAG 0.488 [0.461, 0.515], and RAG+reasoning 0.463 [0.442, 0.487]. The plain and RAG intervals do not overlap, so retrieval's gain is significant (paired bootstrap P(RAG > plain) = 1.000, mean $\Delta = +0.161$). Reasoning, by contrast, is statistically indistinguishable from plain prompting (P = 0.42, mean $\Delta = -0.002$), and adding reasoning to retrieval does not significantly improve it (P = 0.08). The central findings (retrieval helps, reasoning does not) therefore hold at the level of statistical significance, not merely point estimates.

### 7.4 Cost–accuracy frontier

Figure 2 plots micro-F1 against cost per building on a logarithmic axis. The supervised tree sits above all LLM arms at negligible cost; among LLMs, the RAG arms dominate the plain and reasoning arms. The best RAG arm is GPT-5.5 at 0.526, but Llama 3.3 70B with retrieval reaches 0.511 at $0.00032 per building comparable to GPT-5.5 with retrieval (0.526 at $0.00318) at roughly 10.1× lower cost and the lowest latency in the study. Figure 3 shows the precision–recall trade-off by architecture.

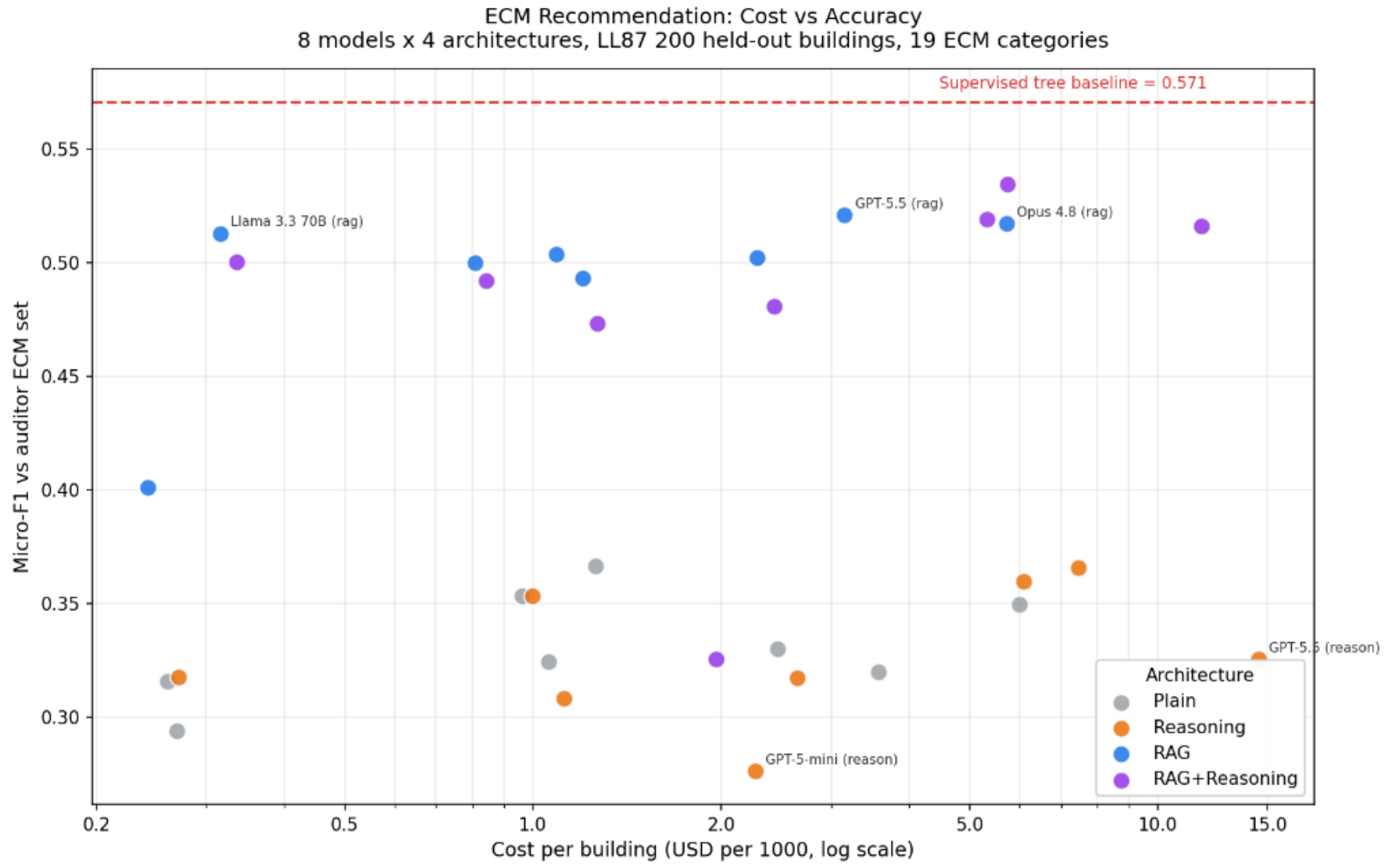

***Figure 2.*** *Cost–accuracy frontier for ECM recommendation. Each point is one model–architecture configuration, plotting micro-F1 against cost per building (log scale; n = 200 held-out buildings). The supervised tree baseline (dashed line, 0.571) exceeds all LLM arms. Among LLMs, the RAG and RAG+reasoning arms (blue, purple) dominate the plain and reasoning arms (grey, orange), and Llama 3.3 70B with retrieval anchors the low-cost frontier at near-frontier accuracy for roughly an order of magnitude less cost.*

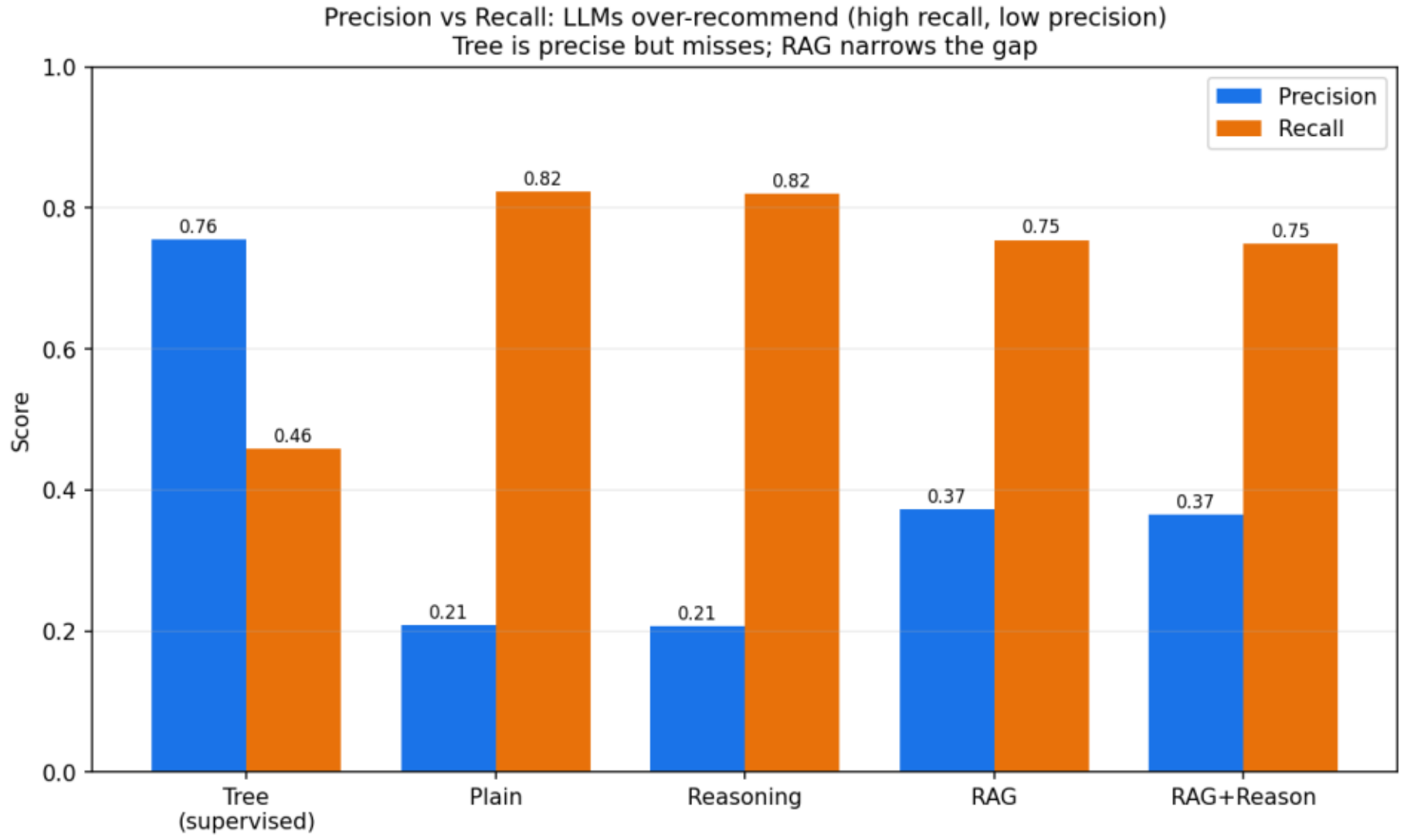

***Figure 3.*** *Precision versus recall by architecture, averaged across the eight LLMs. The supervised tree is precise (0.76) but conservative (0.46 recall), naming few measures and missing many. Plain and reasoning LLMs invert this balance recall near 0.82 but precision near 0.21 systematically over-recommending. Retrieval raises precision to 0.37 while largely preserving recall, indicating that its benefit is the suppression of ill-fitting measures rather than broader coverage.*

# 8. Analysis

## 8.1 Why retrieval helps: precision, not coverage

The precision–recall decomposition (Table 4) reveals the mechanism. The supervised tree is precise at 0.755 but conservative at recall 0.459: it names few measures and misses many true ones. Plain LLMs invert this, with recall near 0.82 and precision near 0.21: they name nearly every category and over-recommend. Retrieval raises LLM precision from roughly 0.21 to roughly 0.37 while largely preserving recall, so its benefit is the suppression of measures that do not fit the building fewer false positives rather than wider coverage. Grounding the model in the real recommendations of similar buildings[26] constrains it toward measures auditors actually selected for comparable stock.

## 8.2 Why reasoning does not help

Across all eight models, enabling reasoning changes micro-F1 by at most 0.018 in either direction (Table 5), while on the small reasoning-heavy models it inflates output tokens by an order of magnitude and cost up to 8.4×. We interpret ECM recommendation as an associative task: the appropriate measures for a building are those recommended for similar buildings, exactly the signal retrieval supplies. Chain-of-thought[11], self-consistency[27], and reason-act scaffolds[28] aid compositional, multi-step problems; they have little to deduce over here in the absence of retrieved evidence, and little to add once it is present. This is a useful negative result for practitioners tempted to pay for reasoning by default.

The token economics make the futility concrete. On GPT-5-mini the reasoning arm emitted roughly 220,000 output tokens across the 200 buildings more than eleven times the plain arm's output for a micro-F1 that fell from 0.295 to 0.276. The additional inference-time compute went into deliberation that did not change, and in several cases worsened, the final category set. This is the signature of a task whose difficulty is informational rather than inferential: the model is not failing for want of reasoning steps but for want of knowledge about which measures auditors attach to buildings like this one. Once retrieval injects that knowledge, the reasoning and plain RAG arms converge to within a hundredth of an F1 point on most models, confirming that reasoning was never the operative variable. The implication is sharp: enabling a reasoning mode by default on this class of task buys nothing and can multiply the bill several-fold.

### 8.3 Where retrieval helps: per-category breakdown

Aggregate F1 conceals strong variation across the 19 categories. Table 7 reports per-category F1 for the supervised tree and for the plain and RAG arms (averaged across the eight models), ordered by category frequency, and Figure 4 visualizes the same. Three regimes emerge. For common measures (Lighting at 56% prevalence, Boiler Plant at 39%), all methods perform reasonably and retrieval adds little: the signal is abundant enough that even an ungrounded model recovers it. For mid-frequency measures, retrieval is decisive: on Renewable Energy Systems the plain LLM scores 0.12 and RAG 0.47, on Distribution Systems 0.29 rises to 0.41, and on Service Hot Water 0.38 rises to 0.45 grounding in similar audited buildings supplies associations the model lacks. For rare measures (at or below 2% prevalence Distributed Generation, Refrigeration, Advanced Metering), the LLM arms collapse to near zero regardless of retrieval, because too few neighbors carry the label to ground on, whereas the supervised tree retains usable F1 (0.44 to 0.67) by learning the category base rates directly. This is the mechanism behind the macro-versus-micro gap: macro-F1 weights every category equally and is therefore dragged down by the rare-category collapse, while micro-F1 is buoyed by the common categories. It also delimits the method's reach: retrieval-grounded LLMs are competitive on common and mid-frequency measures but should defer to a supervised model, or to an auditor, on rare ones.

The frequency threshold at which retrieval stops helping is itself a practically useful quantity. Above roughly 10% prevalence, every retrieval arm improves on its plain counterpart, because a query building's five nearest neighbors are likely to include at least one carrying the measure, so the real auditor recommendation enters the prompt as evidence. Below roughly 2% prevalence, the neighbor set rarely contains the measure at all, and the LLM which only ever sees what retrieval surfaces cannot recover it; its score falls to near zero. The supervised tree does not share this failure mode because it estimates each category's base rate and feature dependence from the full training pool rather than from five neighbors, so it continues to assign the rare measures at their learned frequency. A deployed system should therefore route by prevalence: use the retrieval-grounded LLM for the common and mid-frequency measures that dominate real retrofit packages, and fall back to the supervised model or to human review for the long tail. This hybrid is cheaper than running an auditor on every building and more complete than either model alone.

| ECM category | Freq. | Tree | Plain | RAG | Δ (RAG−Plain) |
|---|---|---|---|---|---|
| Lighting Improvements | 56% | 0.79 | 0.74 | **0.77** | +0.04 |
| Boiler Plant Improvements | 39% | 0.69 | 0.52 | **0.59** | +0.07 |
| Control Systems | 29% | 0.43 | 0.48 | **0.52** | +0.04 |
| Building Envelope Modifications | 28% | 0.52 | 0.36 | **0.43** | +0.07 |
| Chilled Water; Hot Water; and Steam Distribu | 26% | 0.51 | 0.29 | **0.41** | +0.12 |
| Heating; Ventilating and Air Conditioning | 23% | 0.31 | 0.33 | **0.35** | +0.02 |
| Service Hot Water System | 19% | 0.32 | 0.38 | **0.45** | +0.07 |
| Renewable Energy Systems | 19% | 0.54 | 0.12 | **0.47** | +0.35 |
| Water and Sewer Conservation Systems | 14% | 0.46 | 0.33 | **0.44** | +0.11 |
| Electric Motors and Drives | 10% | 0.35 | 0.21 | **0.36** | +0.15 |
| Ventilation System | 6% | 0.46 | 0.09 | **0.18** | +0.09 |
| Appliance and Plug-Load Reductions | 3% | 0.38 | 0.06 | **0.14** | +0.08 |
| Chiller Plant Improvements | 2% | 0.49 | 0.25 | **0.44** | +0.19 |
| Conveyance Systems | 1% | 0.46 | 0.05 | **0.44** | +0.40 |
| Electrical Peak Shaving/Load Shifting | 1% | 0.35 | 0.01 | **0.00** | -0.01 |
| Advanced Metering Systems | 1% | 0.67 | 0.02 | **0.03** | +0.01 |
| Distributed Generation | 1% | 0.56 | 0.00 | **0.00** | +0.00 |
| Energy/Utility Distribution Systems | 1% | 0.47 | 0.00 | **0.00** | +0.00 |
| Refrigeration System Improvements | 1% | 0.44 | 0.00 | **0.00** | +0.00 |

*Table 7. Per-category F1 by frequency. Plain and RAG are averaged across the eight LLMs; Tree is the supervised baseline. Retrieval gains (Δ) are largest for mid-frequency measures; rare measures (bottom rows) collapse for LLMs but hold for the tree.*

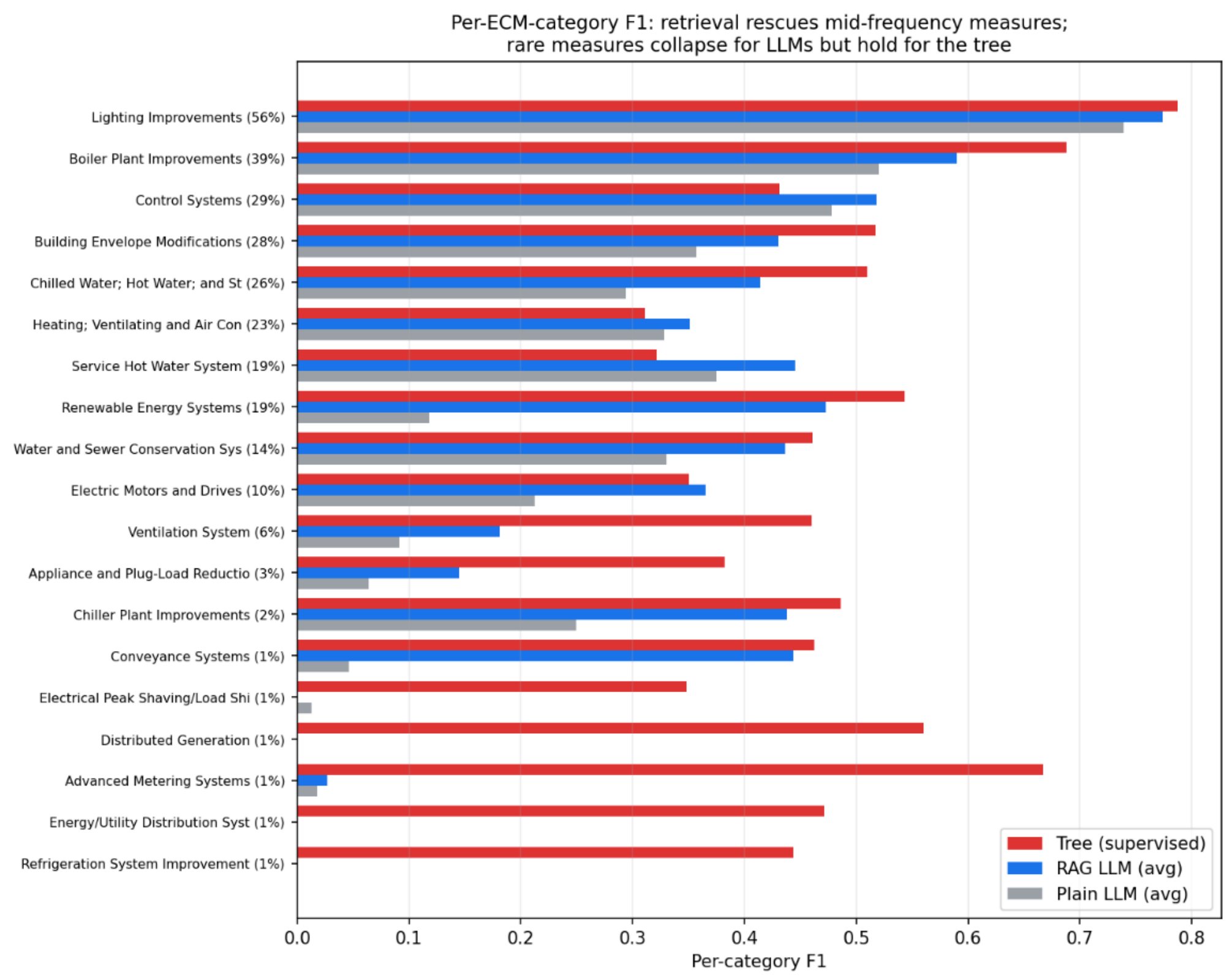

***Figure 4.*** *Per-category F1 across the 19 ECM categories, ordered by prevalence, for the supervised tree and the plain and RAG LLM arms (LLM arms averaged across the eight models). Common measures (top) are handled well by all methods; for mid-frequency measures (e.g., Renewable Energy Systems, Service Hot Water) retrieval produces large gains; for rare measures*

*(bottom, ≤2% prevalence) the LLM arms collapse toward zero while the supervised tree retains usable F1, explaining the gap between macro- and micro-averaged performance.*

### 8.4 Cost decomposition and deployment

Because all eight RAG arms land within a narrow F1 band, the operative deployment decision is cost, not model identity. The open-weight RAG configuration attains near-frontier accuracy at roughly an order of magnitude lower cost and the lowest latency in the study for an operator scoring tens of thousands of buildings, the difference between a feasible and an infeasible triage pipeline. Where labeled data exist, the supervised tree remains both most accurate and cheapest at inference; the LLM's advantage is needing no task-specific training and adapting immediately to a new label space.

### 8.5 Threats to Validity

Several factors bound the strength of these conclusions, and we state them explicitly rather than leave them implicit. The retrieval results depend on the choice of neighbor count $k = 5$; a larger k would surface more candidate measures, likely raising recall and lowering precision, while a smaller k would do the reverse. We did not sweep k, so the reported retrieval numbers are a single operating point, not an optimum; the qualitative finding that retrieval helps and reasoning does not is unlikely to reverse under a different k, but the exact magnitudes would shift. The retrieval index uses Euclidean distance on standardized features, an unweighted metric; a learned or feature-weighted distance could change which neighbors are surfaced and is a natural extension.

The cost comparison rests on provider list prices at a single point in time, and for the two open-weight models on a hosted provider's estimated rates rather than confirmed billing; absolute dollar figures should be treated as indicative, though the order-of-magnitude cost gap between open-weight and frontier RAG is large enough to survive substantial price revision. The accuracy numbers come from a single train/test split at one random seed, so the per-model deltas carry sampling noise on the order of a few F1 points; the consistency of the retrieval effect across all eight models, rather than any single value, is what supports the headline claim. Finally, the corpus is specific to New York City buildings under one regulatory regime, and auditor recommendations reflect both building physics and local program incentives; transfer to other jurisdictions is plausible given the generality of the method but is not demonstrated here.

### 8.6 A Sixth Arm: Direct Generation via Preference Optimization

This paper's conclusion (Section 11) poses a question it does not yet answer: whether a preference-optimized policy trained against auditor recommendations can close the remaining gap to the supervised baseline. We tested this directly. It does not, and the way it fails is informative.

**Goal and setup.** The motivation is product-shaped rather than purely scientific. Section 7 shows RAG (Arm 4) already reaches competitive accuracy, but it depends on a retrieval index and a per-query LLM call at deployment time. A policy that had internalized auditor judgment into its weights would not need either: it could run as a small, self-contained, offline-capable recommendation tool, which is closer to what a paid product needs than a pipeline with external dependencies. Arm 6 tests whether that internalization is achievable at small model scale. Arms 1–5 compare inference-time strategies under an identical input/output contract: every arm receives the same building feature representation and returns a subset of the 19 ECM categories (Section 6). Arm 6 is structurally different. Rather than classify against a fixed label set or retrieve comparable cases, it fine-tunes a small open-weight model, Qwen2.5, via QLoRA [33], to **generate** auditor-quality ECM recommendations directly from a building profile, targeting a standalone, low-cost recommendation product at ≥80% per-building category match. The pipeline combines Direct Preference Optimization (DPO) [34] on auditor-derived preference pairs, a learned reward model, Group Relative Policy Optimization (GRPO) [35] in both learned-reward and ground-truth-reward variants, and rejection-sampling fine-tuning as an alternative to GRPO. Preference pairs are constructed from the same 8,337-building training split used elsewhere in this paper, with the auditor's real recommendation as the preferred response. Because Arm 6 targets generation rather than the Section 6 classification contract, and because its final evaluation uses a held-out set of 20 buildings with a category-match metric rather than the 200-building, per-label-F1 protocol used for Arms 1–5, its results are not directly comparable to Tables 4–7 and are reported separately here.

Figure 5 lays out the pipeline against this target.

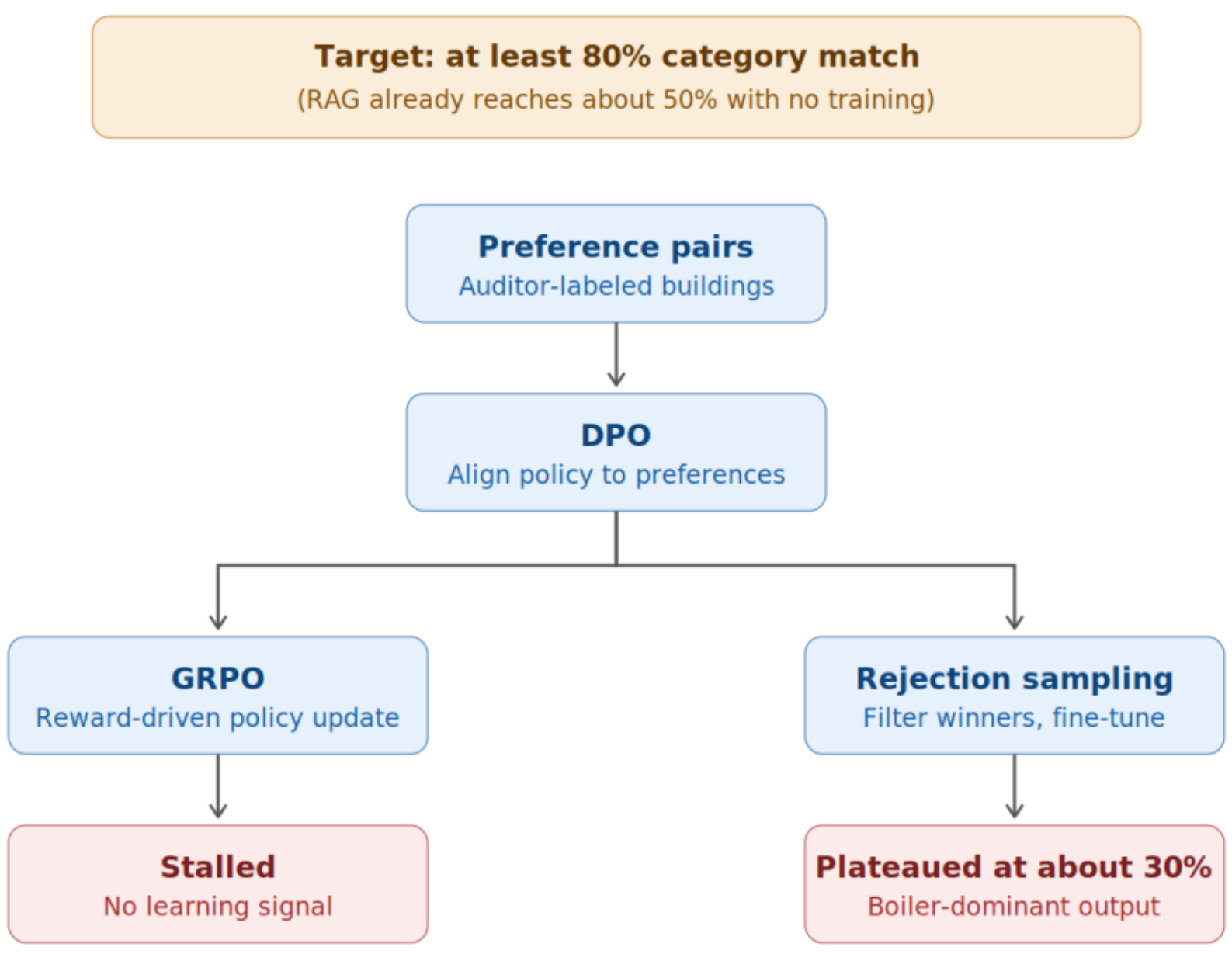

***Figure 5.*** *Arm 6 pipeline and target. The same auditor-derived preference pairs feed a DPO-aligned policy, which branches into GRPO and rejection-sampling fine-tuning. Neither path reaches the 80% target, and both fall short of the roughly 50% the RAG arm already reaches with no training.*

**What was tried.** Table 8 summarizes six progressively larger attempts, from a 1.5B-parameter DPO run on 5,000 preference pairs through a full-dataset run on 21,703 stratified pairs.

| # | Model | Method | Pairs | Result |
|---|---|---|---|---|
| 1 | Qwen2.5-1.5B | DPO | 5,000 | Mode collapse to a single phrase |
| 2 | Qwen2.5-7B | DPO + GRPO, learned reward | 5,000 | clipped_ratio = 1.0; no learning |
| 3 | Qwen2.5-7B | DPO + GRPO, ground-truth reward | 5,000 | Category hit rate flat at ~22% |
| 4 | Qwen2.5-7B | DPO + rejection sampling | 5,000 | 35% category match; boiler-dominant |

| 5 | Qwen2.5-7B | DPO + rejection sampling, stratified (3k cap/category) | 5,000 | 30–35% category match; zero DHW/mechanical/renewable winners |
|---|---|---|---|---|
| **6** | **Qwen2.5-7B** | **DPO + rejection sampling, full-dataset stratified** | **21,703** | **30% category match; no improvement over Attempt 4** |

*Table 8. Arm 6 attempts, in order. All models are Qwen2.5 fine-tuned via QLoRA on a single 8,337-building training split. Category match is measured on a 20-building held-out set and is not the per-label F1 used for Arms 1–5 in Tables 4–7.*

**Why this sequence.** Each escalation in Table 8 targets a specific hypothesis about why the prior attempt fell short, rather than a blind hyperparameter sweep. DPO was the starting point because it needs no separate reward model or RL loop: the cheapest route to pulling a base model toward auditor-preferred outputs (Attempt 1). When the 1.5B policy collapsed to a single repeated phrase, we moved to 7B and added a reward model plus GRPO, on the hypothesis that a learned, denser reward signal would give the policy more to optimize against than a binary DPO preference (Attempt 2); when GRPO's clipped ratio pinned at 1.0 with no learning, we substituted a ground-truth reward computed directly from category match, to remove the learned reward model as a possible confound (Attempt 3). Both GRPO variants underperformed simple rejection sampling (generate many candidates, keep the ones matching the auditor's actual recommendation, fine-tune on the winners), which became the default Phase 2 (Attempt 4). Attempt 4's collapse toward boiler-plant recommendations suggested an input-distribution explanation, so Attempt 5 stratified the preference pairs to cap any single category's share at 13.5%; when rare categories still produced zero rejection-sampling winners, the input-distribution hypothesis was effectively ruled out, and Attempt 6 tested the remaining explanation (insufficient data volume) by scaling the same stratified construction to the full 21,703-pair corpus. That Attempt 6 matched rather than improved on Attempt 4 is itself the result: neither rebalancing the inputs nor multiplying their volume changed what the policy was capable of generating.

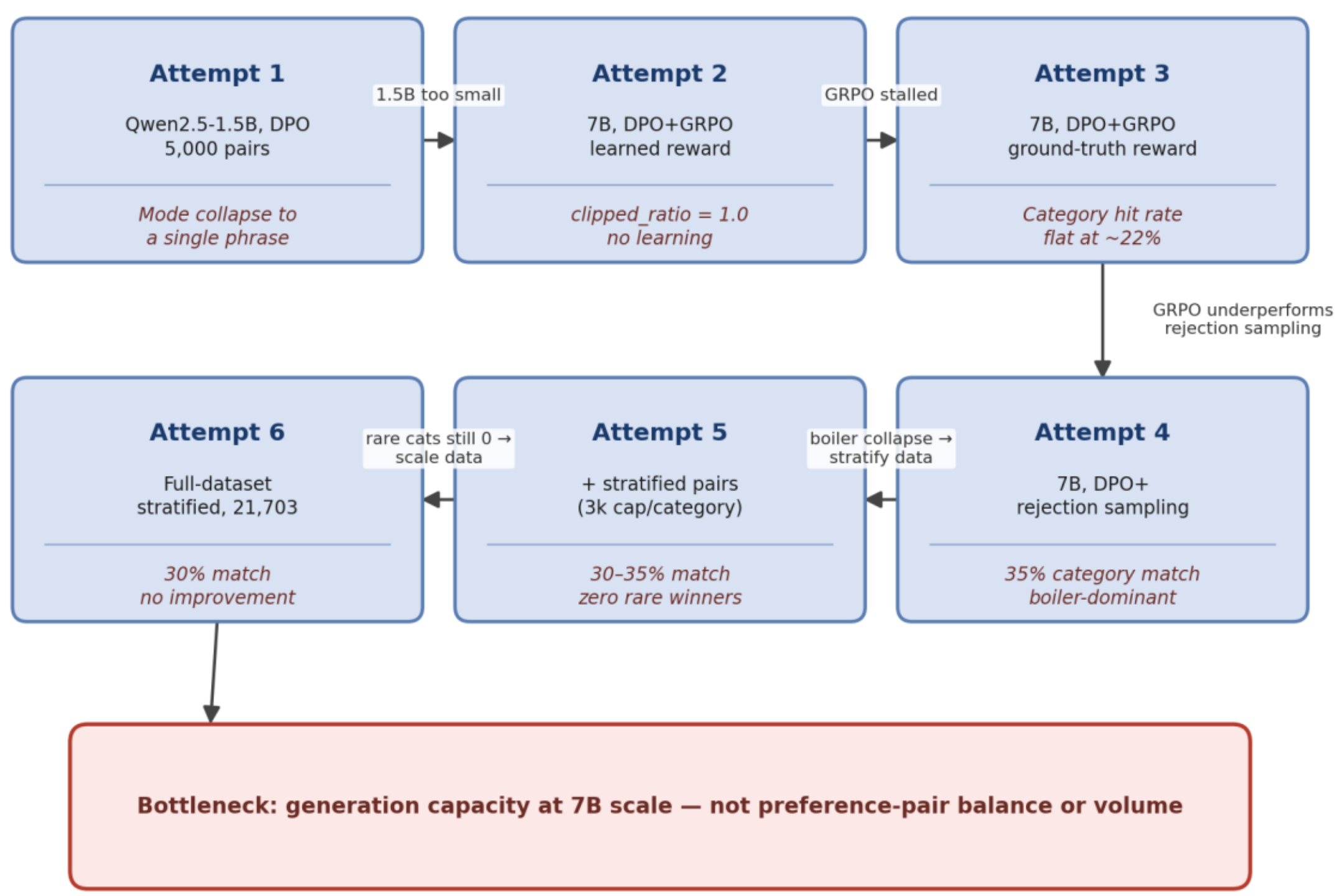

*Figure 6. Arm 6 escalation logic. Each arrow is the specific failure mode of the prior attempt that motivated the next one; the final attempt (full-dataset stratification) tests the data-volume hypothesis and rules it out, isolating generation capacity at 7B scale as the bottleneck.*

**Verifying the saved checkpoint.** We confirmed Attempt 5 directly against its saved adapter rather than against the summary above: its training configuration records **output_dir: rejs_adapter_v2**, matching the asset name in the project's own inventory. The checkpoint is a rank-8, alpha-16 LoRA adapter on the attention projections (q/k/v/o) of all 28 Qwen2.5-7B layers, trained at effective batch size 8 (per-device 2, 4-step gradient accumulation) with a linear schedule from learning rate $2\times10^{-5}$, AdamW, seed 42. The loss recorded at this checkpoint (3.30 at step 10, declining to 3.12 by step 20) sits on the cross-entropy scale appropriate to the rejection-sampling fine-tuning phase described above, not the 0.3–0.7-scale DPO loss reported separately in the next paragraph. This is consistent with the checkpoint capturing Phase 2 (SFT on rejection-sampling winners) rather than the preceding DPO phase.

**Training looked healthy; the outcome did not move.** Considered purely as optimization, the pipeline worked. DPO loss declined cleanly from 0.67 to 0.31 with preference margins growing from 0.05 to 2.3 and training accuracy reaching 0.85; the reward model reached roughly 71% pairwise accuracy; rejection-sampling loss dropped whenever winning candidates existed in a batch. None of this translated into category match. Five materially different attempts (varying model size, reward source, data balance, and data scale by more than 4×) converged on the same 22–35% band, with the best result (Attempt 4, unstratified rejection sampling) matched rather than improved by the largest run (Attempt 6, full-dataset stratified, 21,703 pairs).

**The mechanism: mode collapse, not data scarcity.** Inspection of generations shows why. The policy defaults to recommending boiler-plant upgrades for nearly every building, including ones an auditor would flag for lighting, controls, domestic hot water, or motor and drive measures. NYC's multifamily stock is boiler-heavy, and boiler upgrades became the model's learned safe answer. Across roughly 8,000 candidates generated during rejection sampling, the policy produced zero valid winners in the domestic-hot-water, mechanical, and renewable-energy categories, even after the training pairs were stratified to reduce boiler's share from 27% to 13.5%. Stratification changed the input distribution but not the model's output distribution: a policy that has not learned to generate a category's vocabulary cannot be rejection-sampled into emitting it, however the upstream pairs are balanced. Scaling the stratified pair set from 5,000 to the full 21,703 produced no further movement (Attempt 6 vs. Attempt 5, Table 8). The bottleneck was generation capacity at this model scale, not preference-pair coverage.

**Reading this alongside the rest of the paper.** This negative result is consistent with, and sharpens, the paper's central finding. Sections 7–8 show that retrieval succeeds because it injects auditor knowledge a model does not otherwise have, while reasoning fails because there is nothing to reason over without that knowledge. Arm 6 tests the remaining possibility, that fine-tuning could internalize the missing knowledge into model weights rather than supplying it at inference time, and finds that, at 7B parameters and on the order of 5,000–22,000 preference pairs, it cannot: the model imitates the **format** of an auditor recommendation without acquiring the rare-category vocabulary that auditor judgment depends on. The clean training curves are themselves a useful caution for practitioners: optimizer health (loss, margin, reward-model accuracy) is not evidence of task competence, and reporting it without held-out task accuracy would have been misleading on its own.

We do not take this as evidence that preference optimization cannot work on this task at any scale or data budget; a substantially larger policy, a preference set with deliberate rare-category oversampling beyond the 3× cap used here, or a reward function that explicitly penalizes category collapse are all untested directions. We report it as a negative result bounding the smallest end of that space: a 7B QLoRA policy trained on tens of thousands of auditor-derived preference pairs does not close the gap to the 0.571-F1 supervised baseline, and the RAG arms (Section 7) remain the more practical route to grounding an LLM in this corpus at the model scales we tested.

## 9. Discussion

**Formulation over model.** The largest single improvement came not from any model but from reframing the task as multi-label classification [20] [21], which moved the supervised baseline from a reported 25.9% to 0.571 micro-F1. The honest reading of the LLM results is not that language models surpass supervised learning here they do not but that, given the right formulation, a training-free retrieval pipeline reaches within a few points of a tuned supervised model with no retraining.

**Recursive Language Models, excluded.** Because each prediction concerns a single building whose description fits comfortably in context, the RLM paradigm [29] which earns its advantage on inputs exceeding the context window offers no benefit here. We report this as a tested-and-excluded decision, a negative result useful to practitioners weighing recursive scaffolds for small-input tasks.

**Generalization.** The method is not specific to energy audits: any domain with expert-labeled structured records and a multi-label recommendation target clinical order sets, maintenance work orders, regulatory checklists fits the same template, in which retrieval over expert decisions substitutes for task-specific training. The evaluation design follows holistic, ground-truth-anchored principles [30] [31].

## 10. Limitations

First, results use a single train/test split; a camera-ready version should repeat over multiple seeds and report variance (the harness supports a seed argument). Second, Gemini 3.5 Flash lost 1–32 of 200 evaluations on some arms to transient provider (HTTP 503/504) errors infrastructure, not method and is scored on the completed subset. Third, per-token prices for the two open-weight models are hosted-provider list estimates and should be confirmed against billing before the corresponding dollar figures are quoted as exact. Fourth, the 25.9% single-label figure originates in earlier work on this corpus and is cited as prior context, not produced here. Fifth, the benchmark evaluates ECM-category recommendation only; predicting measure-level savings and payback is a harder, heavy-tailed sub-problem held out by design. Sixth, a per-ECM-category breakdown of LLM performance was not computed because per-building predictions were not logged during evaluation; it can be produced by re-running with prediction logging enabled and is left to future work. Seventh, the Arm 6 preference-optimization experiment (Section 8.6) uses a held-out set of 20 buildings rather than the 200-building set used for Arms 1–5, a category-match metric rather than per-label F1, and a single model family (Qwen2.5, QLoRA) at two parameter scales (1.5B, 7B); its negative result should be read as evidence at that scale and data budget, not as a claim that preference optimization cannot succeed on this task with substantially more compute or data.

## 11. Conclusion

On 10,422 real energy-audit records we separated the contributions of problem formulation, retrieval, and reasoning to structured ECM recommendation, after first establishing that the upstream EUI task is solved by tree ensembles ($R^2$ = 0.757, neural networks trailing). Reformulating recommendation as multi-label classification produced the largest gain, lifting a supervised tree to 0.571 micro-F1. Among LLM approaches, retrieval was decisive improving all eight models by +0.11 to +0.20 while explicit reasoning added no measurable accuracy at up to 8.4× the cost. A cheap open-weight model with a fifteen-line retrieval step matched a frontier model at an order of magnitude lower cost. For practitioners: pose the task as multi-label, ground the model in retrieved real cases, and do not pay for reasoning the task does not use.

Read together, the upstream and downstream results tell one story. In both cases EUI regression and ECM recommendation the binding constraint was the information available to the model, not the model's capacity. Neural networks did not beat trees on EUI because the predictive signal lived in a tabular feature (prior-year intensity) that trees already exploit; reasoning did not beat plain prompting on recommendation because the predictive signal lived in comparable buildings' audits, which only retrieval surfaces. The practical lesson generalizes beyond energy: when a structured-prediction task stalls, the first question is whether the relevant signal is present in the input, not whether the model is large enough. Future work should test this benchmark over multiple random seeds for tighter confidence intervals and extend it to measure-level savings and payback prediction. We also tested whether learning from expert preferences could close the remaining gap to the supervised baseline (Section 8.6): a Qwen2.5 policy trained with DPO, GRPO, and rejection sampling on up to 21,703 auditor-derived preference pairs could not, collapsing toward the majority ECM category rather than acquiring the rare-category vocabulary auditor judgment depends on. At the model scales and data budgets we tested, retrieval (Section 7) remains the more practical route to grounding an LLM in this corpus.

*Reference tracking: entries are numbered in order of first appearance in the text. To rename or re-style a reference, edit only its entry above; in-text markers [n] are positional and remain consistent under renumbering.*

# Appendix A. Glossary of Acronyms

| Acronym | Expansion |
|---|---|
| **AI** | Artificial Intelligence |
| **API** | Application Programming Interface |
| **CNN** | Convolutional Neural Network |
| **CoT** | Chain-of-Thought |
| **DPO** | Direct Preference Optimization |
| **ECM** | Energy Conservation Measure |
| **EUI** | Energy Use Intensity |
| **F1** | Harmonic mean of precision and recall |
| **FN** | False Negative |
| **FP** | False Positive |
| **GBT** | Gradient-Boosted Trees |
| **GRPO** | Group Relative Policy Optimization |
| **HVAC** | Heating, Ventilation, and Air Conditioning |
| **kNN** | k-Nearest Neighbors |
| **LL84** | NYC Local Law 84 (energy benchmarking) |
| **LL87** | NYC Local Law 87 (energy audits) |
| **LLM** | Large Language Model |
| **LSTM** | Long Short-Term Memory |
| **MAE** | Mean Absolute Error |
| **MLP** | Multi-Layer Perceptron |
| **NYSERDA** | New York State Energy Research and Development Authority |
| **QLoRA** | Quantized Low-Rank Adaptation |
| **RAG** | Retrieval-Augmented Generation |
| **RLM** | Recursive Language Model |
| **RMSE** | Root Mean Squared Error |
| **$R^2$** | Coefficient of Determination |
| **SHAP** | SHapley Additive exPlanations |
| **TP** | True Positive |

# Appendix B. Prompt Construction

All LLM arms share a common template. The plain arm presents the building profile and the 19 allowable ECM categories and requests a JSON list of categories. The RAG arms additionally insert, ahead of the target building, the retrieved neighbors' real auditor ECM sets as labeled evidence. The reasoning arms add an instruction to deliberate before answering, or enable the model's native reasoning mode. The output contract is fixed: a JSON array of category strings parsed against the 19-category vocabulary; unparseable outputs fall back to a substring match. Exact prompt strings are provided in the released harness.